\documentclass[a4paper]{article}

\usepackage{amsmath}
\usepackage{amsthm}
\usepackage[linesnumbered,ruled,vlined]{algorithm2e}
\usepackage{tikz}
\usepackage{xspace}
\usepackage{caption}
\usepackage{url}
\usepackage[utf8]{inputenc}
\usetikzlibrary{arrows,decorations.pathmorphing,backgrounds,fit,positioning,shapes.symbols,chains,shapes.geometric,shapes.arrows,calc}

\newcommand{\ie}{\emph{i.e.}}
\newcommand{\eg}{\emph{eg.}}

\title{Blockchain using Proof-of-Interaction\footnote{This paper has been submitted to \emph{FUN with algorithms} conference, with a different title. This is why the introduction may look a little bit informal.}}

\usepackage{authblk}

\author[1,2]{Jean-Philippe Abegg}
\author[1]{Quentin Bramas}
\author[1]{Thomas Noël}

\affil[1]{University of Strasbourg, ICUBE\\
Transchain, Strasbourg}
\affil[2]{Transchain, Strasbourg}

\newcommand{\generatePoI}{\texttt{generatePoI}\xspace}
\newcommand{\checkPoI}{\texttt{checkPoI}\xspace}
\newcommand{\checkMessage}{\texttt{checkMessage}\xspace}
\newcommand{\createServices}{\texttt{createServices}\xspace}
\newcommand{\tourLength}{\texttt{tourLength}\xspace}
\newcommand{\penalties}{\texttt{penalties}\xspace}

\newcommand{\concat}{\cdot}
\newcommand{\ns}{n_S}

\begin{document}
	
	\maketitle
	\begin{center}
jp.abegg@transchain.fr, bramas@unistra.fr, noel@unistra.fr
	\end{center}
	
\begin{abstract}
	This paper we define a new Puzzle called Proof-of-Interaction and we show how it can replace, in the Bitcoin protocol, the Proof-of-Work algorithm.
\end{abstract}

\section{Introduction}

A Blockchain is a data-structure where data can only be appended in blocks. It is maintained in a distributed manner by many participants, who may not trust each-other, and some of which can be faulty or malicious. In order for this data structure to be consistent among the participants, a protocol is used to ensure that every one agrees on the next block that is appended into the Blockchain. The most famous example of such protocol, Bitcoin~\cite{bitcoin}, uses the Proof-of-Work to elect a single participant that is responsible for appending the next block. However, the Proof-of-Work mechanism consumes a lots of computational power~\cite{bitcoin-energy-waste}. 
There have been many attempts to avoid using Proof-of-work based agreement, but each one of them adds other constraints (\eg, small number of nodes, hardware prerequisite, new security threat). 

\textbf{The contribution of this paper is twofold.} First, we propose a better alternative to PoW, called Proof-of-Interaction, which requires negligible computational power. Second, we show how it can be used to create an efficient Blockchain protocol that is resilient against selfish mining, but assumes for now a fixed size network.

The intuition behind our protocol is the following. If we want to avoid doing unnecessary work, we have to ask the expert in this domain: the officials working in French administrations. Our protocol is based on the trivial observation, that any one can relate to, that the French administration is a very complex system, that is unpredictable, slow, and executed by nodes with very small computational power.

Here is a very realistic example of what could happen when a user interacts with the French administration. Assume, for instance, that a user wants to pay his taxes. He contacts a service $S_1$, but is usually forwarded to another service $S_2$ to retrieve a document $D_1$ needed to answer the request. After contacting service $S_2$, the answer will be first to get a specific form from a service $S_3$ and to ask service $S_4$ to sign it, so that he could get document $D_1$. However, the service $S_3$ does not exist anymore, since the last restructuring ($S_2$ was not aware of that), so $S_4$ is contacted without the necessary form and our user actually learns that the whole procedure has changed, because he is now in a specific regime (he is most likely from Strasbourg, where the French laws do not entirely apply). So document $D_2$, from service $S_5$, is needed instead of $D_1$. After a long time, the probability that our user successfully completes his task is very low (he may just stop and wait for the debt recovery to proceed). Interestingly, despite the long time needed to complete the task, no work has been done. This idea is well illustrated in the French cartoon movie ``The Twelve Tasks of Asterix'' where Asterix interacts with the Gaulic Administration\footnote{In french : https://www.youtube.com/watch?v=o2su7cbl-pY}.
Even though the procedure might be very slow and tedious, the verification is very fast. Indeed, the French debt recovery service can detect quickly any deviation from the standard procedure. 

The goal is to use this method to elect a node in the network, in the same way Proof-of-work does. The main idea to understand how the Proof-of-Interaction can be used in a Blockchain context is the following. If a set of users is asked to interact with the French Administration, the first one to complete his task is elected to append a block in the Blockchain.
This paper handles the case where all the participants are known, and we discuss at the end how this assumption could be weakened in future work.

\paragraph*{Related Work}
Proof-of-work~\cite{pow} (PoW) is a method initially intended for preventing spamming attacks. It was then used in the Bitcoin protocol~\cite{bitcoin} as a way to prove that a certain amount of time has passed between two consecutive blocks. Another way to see the aim of the Proof-of-work in the Bitcoin protocol is as a leader election mechanism, to select who is responsible for writing the next block in the blockchain. This leader election has several important properties, including protection against Sybil attacks~\cite{sybil-attacks} and against denial-of-service~\cite{dos-attacks}. Also, it has a small communication complexity. However, the computational race consumes a lot of energy~\cite{bitcoin-energy-waste}, that some people might consider wasted (the meaning of life and of the universe is outside the scope of this paper). 

In 2012, S. King and S. Nadal~\cite{king2012ppcoin} proposed the Proof-of-Stake (PoS), an alternative for PoW. This leader election mechanism requires less computational power but has security issues~\cite{stake_bleeding} (\eg, Long range attack and DoS). Intel proposed another alternative to PoW, the Proof-of-Elapsed-Time (PoET)~\cite{poet}. This solution requires Intel SGX as a trusted execution environment. Thus, Intel becomes a required trusted party to make the consensus work, which might imply security concerns~\cite{PoET_security} and is against blockchain idea to remove third parties.

The previous paper most related to our work was presented just prior the presentation of Bitcoin in 2008, by M. Abliz and T. Znati~\cite{guided-tour}. They proposed \emph{A Guided Tour Puzzle for Denial of Service Prevention}, which is another spam protection algorithm. This mechanism has not yet been used in the Blockchain context, and is at the core of our new Proof-of-Interaction. The idea was that, when a user wants to access a resource in a server that is heavily requested, the server can ask the user to perform a tour of a given length in the network. This tour consists of accessing randomly a list of nodes, own by the same provider as the server. After the tour, a user can prove to the server that it has completed the task and can then retrieve the resource. The way we generate our tour in the French administration is based on the same idea. We generalized the approach of M. Abliz and T. Znati to work with multiple participants, and we made the tour length variable.

\vspace{-0.4cm}
\paragraph*{Model}

\newcommand{\N}{\ensuremath{\mathcal{N}}}
The network, is a set $\N$ of $n$ nodes that are completely connected. Each node has a pair of private and public cryptographic keys. Nodes are uniquely identified by their public keys (\ie, the association between the public keys and the nodes is common knowledge). Each message is signed by its sender, and a node cannot fake a message signed by another (non-faulty) node.

\newcommand{\sign}[1]{\ensuremath{\texttt{sign}_{#1}}}
\newcommand{\verif}[1]{\ensuremath{\texttt{verif}_{#1}}}
We denote by $\sign{u}(m)$ the signature by node $u$ of the message $m$, and $\verif{u}(s,m)$ the predicate that is true if and only if $s = \sign{u}(m)$. For now, we assume the signature algorithm is a deterministic one-way function that depends only on the message $m$ and on the private key of $u$. This assumption might be very strong as, with common signature schemes, different signature could be generated for the same message, but there are ways to remove this assumption by using complex secret generation and disclosure schemes, not discussed in this paper, so that each signature is in fact a deterministic one-way function.
\newcommand{\Hsize}{\ensuremath{H_\mathit{size}}}
\newcommand{\Hlen}{\ensuremath{H_\mathit{len}}}
The function $H$ is a cryptographic, one-way and collision resistant, hash function~\cite{hash-functions}. 

As for the Bitcoin protocol, we assume the communication is partially synchronous \ie, there is a fixed, but unknown, upper bound $\Delta$ on the time for messages to be delivered.

The size of a set $S$ is denoted with $|S|$.

\vspace{-0.4cm}
\paragraph*{Naive Approach}
We give here a naive approach on how asking participants to perform a tour in the network can be used as a leader election mechanism to elect the node responsible for appending the next block in a Blockchain.

When a node $u_0$ wants to append a block to the blockchain, he performs a random tour of length $L$ in the network retrieving signatures of each participants he visits. The first node $u_1$ to visit is the hash of the last block $h_0 = last\_block\_hash$ of the blockchain modulo $n$ (if we order nodes by their public keys, the node to visit is the $i$-th with $i= h_0 \mod n$). $u_1$ responds with the signature $s_1$ of $h_0$. The hash $h_1 = H(s_1)$, modulo $n$, gives the second node $u_2$ to visit, and so on. This idea proposed by M. Abliz and T. Znati~\cite{guided-tour} is interesting because the whole tour depends only on the initial value, here the hash of the last block\footnote{in the original article~\cite{guided-tour}, the initial value was a random value signed by the server}. Given $h_0$, any one can verify that the sequence of signatures $(s_1, s_2, \ldots, s_L)$ is a proof that the tour has been properly performed. If each node in the network performs a tour, the first node to complete its tour broadcast its block to the other nodes to announce it.

However, here, each node has to perform the same tour, which could be problematic. An easy fix is to select the first node to visit, not directly using the hash of the last block, but based on the signature of the node initiating the tour, $h_0 = H(\sign{u_0}(last\_block\_hash))$. Now, given $h_0$, the sequence of signatures $(\sign{u_0}(h_0), s_1, s_2, \ldots, s_L)$ proves that the tour has been properly performed by node $u_0$. Each tour, performed by a given node, is unique, and a node cannot compute the sequence of signature other than by actually asking each node in the tour to sign a message. Indeed, the next hop of the tour depends on the current one.

Here, one can see that it could be a good idea to also make the tour dependent on the content of the block node $u_0$ is trying to append. Indeed, using only the last block to generate a new proof does not protect the content of the current block, \ie, the same proof can be used to create two different blocks. To prevent this behavior, we can assume that $h_0 = H(\sign{u_0}(last\_block\_hash) \concat M)$ ($\concat$ being the concatenation operator) where $M$ is a hash of the content of the block node $u_0$ is trying to append. In practice, it is the root of the Merkel tree containing all the transactions of the block. Here, the proof is dependent on the content of the block, which means that if the content of the block changes the whole proof needs to be computed again.

From there, we face another issue. Each node performs a tour of length $L$, so each participant will be elected roughly at the same time, creating a lots of forks. To avoid this, we can make the tour length variable. We found two ways to do so. The first one is not to decide on a length in advance, and perform the tour until the hash of $k$-th signature is smaller than a given target value, representing the difficulty of the proof. In this way, every interaction with another node during a tour can be seen as a tentative to find a good hash (like hashing a block with a given nonce in the PoW protocol). The target value can be selected so that the average length of the tour is predetermined. However doing so, since the proof does depend on the content of the block, $u_0$ can change the content of the block, by adding dummy transactions for instance, so that the tour stops after one hop\footnote{There are some ways to limit this attack, but we believe it will remain an important attack vector}.
The other way to make the tour length variable is to use a cryptographic random number generator, seeded with $\sign{u_0}(last\_block\_hash)$, to generate the length $L$. Doing so, the length depends only on $u_0$ and on the previous block. Then a tour of length $L$ is performed as usual.

To complete the scheme, we add other information to the message sent to the visited node so that they can detect if we try to prove different blocks in parallel. We also make $u_0$ sign each response before computing the next hop, so that the tour must pass through $u_0$ after each visit. 
Finally, we will see why it is important to perform the tour, not using the entire network, but only a subset of it. 

\noindent\textbf{Paper Structure.}
The next section describes our final approach of the Proof-of-Interaction, that could be used outside of the Blockchain context. Then, in Section~\ref{sec:blockchain} we explain how we use this proof mechanism to create a Blockchain protocol.
Then, Section~\ref{sec:security} shows the security properties of our protocol. Finally, we conclude and discuss possible extensions in Section~\ref{sec:conclusion}.

\vspace{-0.2cm}
\section{The Proof-of-Interaction}
\vspace{-0.1cm}

In this section we define the most important piece of our protocol, which is, how a given node of the network generate a proof of interaction. Then, we will see in the next section how this proof can be used as an election mechanism in our Blockchain protocol.

\vspace{-0.2cm}
\subsection{Algorithm Overview}
\vspace{-0.1cm}

We present here a couple of algorithms. One that generates a Proof-of-Interaction (PoI), and one that checks the validity of a given PoI.

\vspace{-0.4cm}
\paragraph*{Generating a Proof-of-Interaction}
\vspace{-0.3cm}

Consider we are a node $u_0\in\N$ that wants to generate a PoI. Given a fixed \emph{dependency} value denoted $d$, the user $u_0$ wants to prove a \emph{message} denoted $m$. The user has no control over $d$ but can chose any message to prove.

\begin{figure}
\begin{minipage}{0.48\textwidth}
    \centering
    \includegraphics[width=5.8cm]{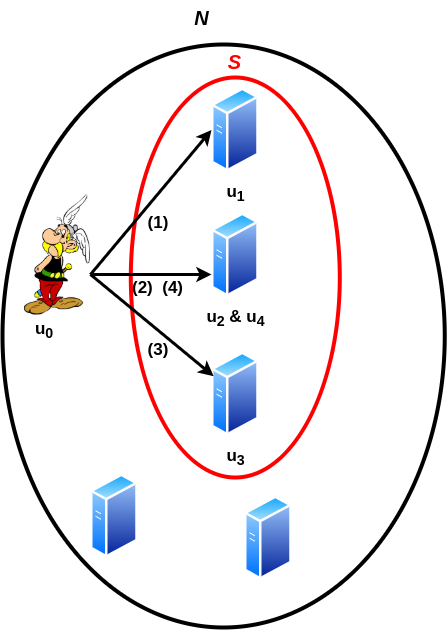}
    \captionof{figure}{Asterix interacts randomly with a subset of the nodes}
    \label{fig:generating a PoI, network view}
\end{minipage}
\begin{minipage}{0.48\textwidth}
    \centering
    \includegraphics[width=5.8cm]{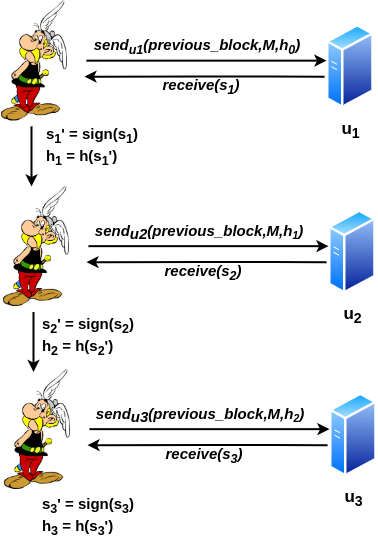}
    \captionof{figure}{Asterix interacts with a sequence of nodes to construct a PoI, In this example, the dependency is the hash of the last block.}
    \label{fig:generating a PoI}
\end{minipage}
\end{figure}

The signature by $u_0$ of the dependency $d$, denoted $s_0 = \sign{u_0}(d)$, is used to generate the subset $S$ of $\ns=\min(20,n/2)$ nodes to interact with, $S = \{S_0, S_2, \ldots, S_{\ns-1}\} = \createServices(\N, s_0)$. $S$ is generated using a pseudo-random procedure, and depends only on $d$ and on $u_0$.
From $s_0$, we also derive the length of the tour $L = \tourLength(D, s_0)$, where $D$ is a probabilistic distribution that corresponds to the difficulty parameter. $\tourLength$ is a random number generator, seeded with $s_0$, with generating a number according to $D$. Using $D$ one can easily change the average length of the tour for instance.

Now $u_0$ has to visit randomly $L$ nodes in $S$ to complete the proof, as illustrated in Figure~\ref{fig:generating a PoI, network view}. To know what is the first node $u_1$ we have to visit, we first hash the concatenation of $s_0$ with $m$ to obtain $h_0 = H(s_0\concat m)$. This hash (modulo $|S|$) gives the index $i$ in $S$ of the node we have to visit, $i = h_0 \mod |S|$. So we send the tuple $(h_0, d, m)$ to node $u_1 = S_i$, which responds by signing the concatenation, $s_1 = \sign{u_1}(h_0 \concat d \concat m)$. 

To know what is the second node $u_2$ we have to visit, we sign and hash the response from $u_1$ to obtain $h_1 = H(\sign{u_0}(s_1))$, so that $u_2 = S_j \in S$ with $j = h_1 \mod  |S|$. Again, we send the tuple $(h_1, d, m)$ to $u_2$, which responds by signing the concatenation, $s_2 = \sign{u_2}(h_1 \concat d \concat m)$. 
We sign and hash the response from $u_1$ to obtain $h_2 = H(\sign{u_0}(s_2))$ and find the next node we have to visit, and so on (see Figure~\ref{fig:generating a PoI}).
This continues until we compute $\sign{u_0}(s_L)$, after the response of the $L$-th visited node.

The \emph{Proof of Interaction} (PoI) with dependency $d$ of message $m$ by node $u_0$ and difficulty $D$ is the sequence 
\[
(s_0, s_1, \sign{u_0}(s_1), s_2, \sign{u_0}(s_2), \ldots, s_L, \sign{u_0}(s_L)).
\]

\vspace{-0.2cm}
\paragraph*{Checking a Proof-of-Interaction}

To check if a PoI $(s_0, s_1, s_1' \ldots, s_k, s_k')$ from user $u$, is valid for message $m$, dependency $d$ and difficulty $D$, one can first check if $s_0$ is a valid signature of $d$ by $u_0$. If so, then we can obtain the set $S=\createServices(\N, s_0)$ of interacting nodes, the length $L=\tourLength(D, s_0)$, and the hash $h_0 = H(s_0\concat m)$. From $h_0$ and $S$, we can compute what is the first node $u_1$ and check if $s_1$ is a valid signature from $u_1$ of $(h_0 \concat d \concat m)$, and if $s_1'$ is a valid signature of $s_1$ from $u_0$. Similarly, one can check all the signatures until $s_k'$. Finally, if all signatures are valid, and $k=L$, the PoI is valid.

\subsection{Algorithm Details}

The pseudo code of our algorithms are given in the appendix.

The algorithm \createServices is straightforward. We assume that we have a random number generator (RNG) that we initialize with the given seed. The algorithm then shuffles the input array using the given random number generator. Finally, it simply returns the first $\ns$ elements of the shuffled array.


The main part of the algorithm \generatePoI consists in a loop, that performs the $L$ interactions. The algorithm requires that each node in the network is executing the same algorithm (it can tolerates some faulty nodes, as explained later). The end of the algorithm shows what is executed when a node receives a message from another node. The procedure $\checkMessage$ may depends on what the PoI is used for. In our context, the procedure checks that the nodes that interacts with us does not try to create multiple PoI with different messages, and use the same dependency as everyone else. We will see in details in the next section why it is important.

The algorithm \checkPoI that checks the validity of a PoI is checking that each signature from the proof is valid and respects the proof generation algorithm.
 
\vspace{-0.4cm}
\paragraph*{Proof-of-Interactions Properties}

Now we explain that the Proof-of-Interaction has the several properties that are awaited by good client-puzzle protocols~\cite{non-parallelizable-client-puzzle}.

\noindent\textbf{Computation guarantee:} The proof can only be generated by making each visited node sign a particular message in the correct order. The sequence of visited node depends only on the initiator node, on the dependency $d$, and on the message $m$, and cannot be known before completing the tour.  Furthermore, a node knows the size of his tour before completing it. Which means that he knows before doing his tour how much messages he needs to exchange and how much signatures he will do to have a correct proof.

\noindent\textbf{Non-parallelizability:} A node cannot compute a valid PoI for a given dependency $d$ and message $m$ in parallel. Indeed, In order to know what is the node of the $i$-th interaction, we need to know $h_{i-1}$, hence we need to know $s_{i-1}$. $s_{i-1}$ is a signature from $u_{i-1}$. So we can interact with $u_i$ only after we receive the answer from $u_{i-1}$ \ie, interactions are sequential.

\noindent\textbf{Granularity:} The difficulty of our protocol is easily adjustable using the parameter $D$. The expected time to complete the proof is $2 \times \mathit{mean}(D) \times \mathit{Com}$ where $\mathit{Com}$ is the average duration of a message transmission in the network, and $\mathit{mean}(D)$ is the mean of the distribution $D$.

\noindent\textbf{Efficiency:} Our solution is efficient in terms of computation for all the participants. The generation of one PoI by one participant requires $\mathit{mean}(D)$ hashes and $\mathit{mean}(D)$ signatures in average for the initiator of the proof, and $\mathit{mean}(D)/n$ signatures in average for another node in the network. The verification requires $2D+1$ signature verification and $\mathit{mean}(D)$ hashes in average. The size of the proof is also linear in the difficulty, as it contains $2\mathit{mean}(D)+1$ signatures.

\section{Blockchain Consensus Using PoI}\label{sec:blockchain}

In this section we detail how we can use the PoI mechanism to build a Blockchain protocol. The main idea is to replace, in the Bitcoin protocol, the Proof-of-work by the Proof-of-interactions, with some adjustments. We prove in the next section that it provides similar guarantees to the Bitcoin protocol. 

\newcommand{\mypar}[1]{\paragraph*{#1}}
\vspace{-0.1cm}
\mypar{Block Format}
First, like in the Bitcoin protocol, transactions are stored in blocks that are chained together by including in each block, a field containing the hash of the previous block. In Bitcoin, a block includes a nonce field so that the hash of the block is smaller than a target value (hence proving that computational power has been used) whereas in our protocol, the block includes a proof of interaction where the dependency $d$ is the hash of the previous block, and the message $m$ is the root of the Merkel tree storing the transactions of the current block. Like for the transactions, the block header could contains only the hash of the PoI, like illustrated in Figure~\ref{fig:block} (in the appendix), and the full proof can be stored in the block data, along with the sequence of transactions.

\vspace{-0.1cm}
\mypar{Block Generation}
How is the next block appended in the blockchain? Like in Bitcoin, each participant has gathered a set of transactions (not necessarily the same) and wants to append a block to the blockchain. Do to so, each one of them tries to generate a PoI with the hash of the last block as dependency $d$, the root of the Merkel tree of the transactions of their own block as message $m$, and using the last block difficulty $D$. We assume the difficulty $D$ is characterized by its mean value $\mathit{mean}(D)$, which is the number that is stored into the block.

Participants have no choice over $d$ so the length of their tour, and the subset $S$ of potential visited node is fixed for each participants (one can assume that it is a random subset). Each participant is trying to complete its PoI the fastest as possible, and the first one that completes it, has a valid block. The valid block is broadcasted into the network to announce to everyone that they now have complete a PoI using this new block as dependency.
When a node receives a block from another node, it checks if all the transactions are valid and the checks if the proof of interaction is valid. If so, it appends the new block to its local blockchain and starts generating a proof of interaction based on this new block. 

First, one can see that this could lead to forks, exactly like in the Bitcoin protocol, where different part of the networks try to generate PoI with different dependencies. Thus, the protocol dictates that only one of the longest chain should be used as a dependency to generate a PoI.
This is defined in the procedure \checkMessage. When a node receives a message from another node, it first checks if the dependency matches the latest block of one of the longest chain. If not, the request is ignored.

\vspace{-0.2cm}
\mypar{Incentives}
Like in Bitcoin, we give incentives to nodes that participate to the protocol. The block reward (that could be fixed, decreasing over time, or just contains the transactions fees) is evenly distributed among all the participants of the PoI of the block. This implies that, to maximize their gain, nodes should answer as fast as possible to all the requests from the other nodes currently generating their PoI.

Also, it means that we do not want to answer a request for a node that is not up to date \ie, that is generating a PoI for a block for which there is already a valid block on top, or for a block in a branch that is smaller than longest one. 

\vspace{-0.2cm}
\mypar{Preventing Double-Touring Attacks}
What prevents a node to try to generate several PoI using different variation of its block? If a node wants to maximize its gain (without even being malicious, just rational) it can add dummy transactions to its current block to create several versions of it. Each version can be used to initiate the generation of a PoI using different tours. However, he has to send the message $m$ every times he interacts with another node. If the length of the tour is long enough, the probability that two different tours intersect is not negligible. In other words, a node that receive two messages from the same node, but with different values of $m$ will raise the alarm. To prevent \emph{double-touring}, it is easy to add an incentive to discourage nodes from generating several blocks linked to the same dependency. To do so, we assume each participant has locked a certain amount of money in the Blockchain, and if a node $u$ has a proof that another node has created two different blocks with the same dependency (\ie, previous block), then the node $u$ can claim as reward the locked funds of the cheating node. In addition, it can have other implications such as the exclusion of the network. We assume that the potential loss of being captured is greater than the gain (here the only gain would be to have a greater probability to append its own block).

\vspace{-0.2cm}
\mypar{Difficulty Adjustment}
The difficulty could be adjusted exactly like in Bitcoin. The goal is to chose the difficulty so that the average time $B$ to generate a block is fixed. Here, the difficulty parameter $D$ gives a very precise way to obtain a delay $B$ between blocks and to limit the probability of fork at the same time.
If $\mathit{Com}$ denotes the average duration of a transmission in the network, then we want the expected shortest tour length among the participants to be $\lceil B/\mathit{Com}\rceil$.

For instance, if $D$ is the uniform distribution between $1$ and $\lceil B/\mathit{Com}\rceil(n+1)-1$, then the length of the shortest tour length among all the participants will be exactly $\lceil B/\mathit{Com}\rceil$.

Every given period (\eg, 2016 blocks as in Bitcoin), the difficulty could be adjusted using the duration of the last period (using the timestamps included in each block) to take into account the possible variation of $\mathit{Com}$, so that the average time to generate a block remains~$B$.



\vspace{-0.2cm}
\mypar{Communication Complexity}
A quick analysis shows that each node sends messages sequentially, one after receiving the answer of the other. At the same time, it answers to signature requests from the other nodes. In average, a node is part of $\ns$ tours. Hence the average number of messages per unit of time is fixed \ie, $\ns+1$ every $\mathit{Com}$. Then, the total amount of messages, per unit of time, in the whole network is linear in $n$.

\section{Security}\label{sec:security}

This section discusses about common security threats and how our PoI-based Blockchain handles them. We assume that honest nodes will always follow our algorithms but an attacker can have arbitrary behavior, while avoiding receiving any penalty (which could remove him from the network). We assume that an attacker can eavesdrop every messages exchange between two nodes but he can not change them. Also, assume that an attacker $A$ can not forge messages from another honest node $B$.

\vspace*{-0.2cm}
\paragraph*{Crash Faults}

A node crashes when it completely stops its execution. The main impact is that it does not respond to the sign requests of other nodes. This can be an issue because at each step of the PoI generation, the initiator node could wait forever the response of a crashed node. 
Crashed nodes are handle by the fact that a node only has to interact with a subset $S$ of the whole network $\N$, computed using the service creation function, $\createServices$. 
Hence, if a node crashes, only a fraction of the PoI that are being generated will be stuck waiting for it. All the nodes whose Service set $S$ do not contains crash faults are able to generate their PoI entirely. Since each set $S$ is of size $\ns = \min(n/2, 20)$, we have that, if half of the nodes crash, the probability a given set $S$ contains a crashed node is $1-\left(\frac{1}{2}\right)^{n_S}$. So that the probability $p$ that at least one set $S$ contains only correct nodes is
\[
p = 1-\left(1-\left(\frac{1}{2}\right)^{n_S}\right)^n
\]
One can see that the probability $p$ tends quickly (exponentially fast) to 1 as $n$ tends to infinity. For small values of $n$, the probability is greater than a fixed non-null value. In the rare event that all the sets $S$ contain at least a crashed node, then the protocol is stuck until some crashed nodes reboot and are accessible again.

Finally, we recall that honest nodes are incentivized to answer, because they get a reward when they are included in the next block's PoI. Hence, honest nodes will try be back again as fast as possible.

\vspace*{-0.2cm}
\paragraph*{Selfish mining}

Selfish mining~\cite{selfish-mining} is an attack where a set of malicious nodes collude to waste honest nodes resources and get more reward. It works as follow. Once a malicious node finds a new block, he only shares it with the other malicious nodes. All malicious nodes will be working on a private chain without revealing their new block, so that honest nodes are working on a smaller public branch \ie, honest nodes are wasting resources to find blocks on a useless branch. When honest nodes find a block, the malicious nodes might reveal some of it private blocks to discard honest blocks and get the rewards.

In Bitcoin, selfish mining is a real concern as attackers having any fraction of the whole computational power could successfully use this strategy~\cite{optimal-selfish-mining}.

Interestingly, our PoI-based Blockchain is less sensible to such attacks.
Our algorithm gives a protection by design. Indeed, when generating a PoI, a node has to ask to a lots of other nodes to sign messages containing the hash of the previous block, forcing it to reveal any private block. Other nodes in the network will request the missing block before accepting to sign the message. In other words, it is not possible to generate a PoI alone. Moreover, if a node is working on a branch that is smaller than the legitimate chain and ask for the signature of an honest node, the latter will tell the former to update its local Blockchain, thus preventing him from wasting resources.

\vspace*{-0.2cm}
\paragraph*{Shared Mining}
During the PoI, a node will most of the time be waiting for the signature of another node. So the network delay has the highest impact on the block creation time. To remove this delay, a set of malicious nodes can share theirs private keys between each other and try to create a set $S$ where every nodes are malicious. If one malicious node of the pool succeeds, he can compute the proof locally without sending any messages. He will generate the PoI faster than honest nodes and have a high chance to win. 

We defined earlier that each node of the network is known. Which mean that each node is a distinct entity. For this attack to succeed, entities need to share their private keys. This is a very risky move because once you give your private key to someone, he can create transactions in your name without your authorization. This risk alone should discourage honest nodes to do it, even if they want to maximize their gain. 

We can still assume that a small number of malicious nodes do know each other and collude to perform this attack. We show know that this attack is hard to perform. $S$ only depends on the previous block and on the identity of the initiator of the proof, so the nodes have no control over it. $S$ consists of $n_S$ nodes randomly selected among the network. So if there are $F$ malicious friends on the network, there is on average the same fraction $(n/F)$ of malicious friends in $S$ as in $\N$.
However, the probability for the tour to contains only his friend is very low. Indeed, with $F$ malicious friends on the network, the probability that the entire tour consists of malicious friends is $(n/F)^{\mathit{mean}(D)}$ in average. 


When a malicious node initiates a PoI for a given message, it can see whether the tour contains an honest node or not, 
so he might be tempted to change the content (by reordering the transaction or inserting dummy transactions) of its block until the tour contains only his malicious friends.
However, if there is a fraction $(n/F) = 0.1$ of malicious friends in the network (hence in $S$), and if $\mathit{mean}(D) = 100$, for instance, then the probability that a given tour contains only malicious friends is $10^{-100}$. To find a tour with only malicious friends, an initiator would have to try in average $10^{100}$ different block content, which is not feasible in practice.

\section{Conclusion and Possible Extensions}\label{sec:conclusion}

We have presented an new puzzle mechanism that requires negligible work from all the participants. It requires participants to gather sequentially a list of signatures from a subset of the network, forcing them to wait for the response of each visited node. This mechanism can be easily integrated into a Blockchain protocol, by replacing the energy inefficient Proof-of-work.
The resulting Blockchain protocol is efficient and more secure than the Bitcoin protocol as it is not subject to selfish-mining. Also, it does not have the security issues found in usual PoW replacements such as Proof-of-stack or Proof-of-elapsed time. However, it currently works only in networks where participants are known in advance. The design of our Blockchain protocol makes it easy to propose a possible extension to remove this assumption. 

The easiest way to allow anyone to be able to create blocks, is to select as participants, the $n$ nodes that locked the highest amount of money. This technique is similar to several existing blockchain based on protocols that work only with known participants (such as Tendermint~\cite{tendermint} using an extension of PBFT~\cite{PBFT}) or where the nodes producing blocks are reduced for performance reasons (such as EOS~\cite{eos} where 21 producer nodes are elected by votes from stakehodlers).
We believe a vote mechanism from stakeholders can elect the set of participants executing our protocol. The main advantage with our solution is that the number of participants can be very high, especially compared to previously mentionned protocols.



\bibliography{bib}
\newpage
\appendix
\section{Omitted Algorithm}

\begin{algorithm}[t]
\DontPrintSemicolon
\KwIn{
	$d$, the dependency (hash of last block of the blockchain) \\
	$m$, the message (root of the merkle tree of the new block) \\
	$D$, difficulty of the PoI\\
	$N$, the set of nodes in the network \\
}
\KwOut{
	$P$, a list of signatures $\{s_0,s_1,s_1', s_2, s_2', \ldots, s_k, s_k' \}$ \\
}
	$P \gets [\,]$\;
	$s_0 \gets \sign{u_0}(d)$\;
	$S \gets \createServices(N,s_0)$\;
	$L \gets \tourLength(D,s_0)$\;
	$P.append(s_0)$\;
	$current\_hash \gets H(s_0 \concat m)$\;
    \For{$L$ iterations}{
        $next\_hop \gets current\_hash\%|S|$\;
        $ s \gets send_{S_{next\_hop}}(current\_hash, d, m)$\;
        $ P.append(s)$\;
        $ s \gets  \sign{u_0}(s) $ \;
  		$ P.append(s)$\;
        $ current\_hash \gets H(s)$\;
    }
	\Return{$P$}\;
	\vspace{0.4cm}
	\textbf{When Receive} $(h,d,m)$ from $u$ \textbf{do}

    \If{$\checkMessage(u,h,d,m)$}{
        \textbf{Reply} $\sign{u_0}(h \concat d \concat m)$
    }
\SetAlgoRefName{{\tt generatePoI}}
\caption{Program executed by $u_0$ to generate the PoI}\label{algo:pofai}
\end{algorithm}

\begin{algorithm}[t]
\SetAlgoRefName{{\tt checkPoI}}
\caption{Program executed by anyone to check the validity of a PoI}
\label{algo:pofai_verif}
\KwIn{
	$P$, a proof-of-interaction \\
	$u$, creator of the proof \\
	$d$, the dependency (hash of last block of the blockchain) \\
	$m$, the message (root of the merkle tree of the new block) \\
	$D$, difficulty of the PoI\\
	$N$, the set of nodes in the network
}
\KwOut{
	whether $P$ is a valid PoI or not
}
    \If{not \verif{u}(P[0],d)}{
        \Return{$false$}
    }
    $S \gets \createServices(N,P[0])$\;
	$L \gets \tourLength(D,P[0])$\;
	\If{$L*2 + 1 \neq |P|$}{
           \Return{$false$}
    }
	$current\_hash \gets H(P[0] \concat m)$\;
	\For{$i = 0;\ i < L;\ i++$}{
        $next\_hop \gets current\_hash\%|S|$\;
        \If{not $\verif{S_{next\_hop}}(P[2*i+1], current\_hash \concat d \concat m)$}{
           \Return{$false$}
        }
         \If{not $\verif{u}(P[2*i+2],P[2*i+1])$}{
            \Return{$false$}
        }
        $ current\_hash \gets H(P[2*i+2])$\;
    }

	\Return{$true$}
\end{algorithm}

\begin{algorithm}[t]
\SetAlgoRefName{{\tt createServices}}
\caption{create a pseudo-random subset of nodes}
\label{algo:createServices}
\DontPrintSemicolon
\KwIn{
	$N$, the set of nodes \\
	$h$, a seed
}
\KwOut{
	$S$, a subset of nodes
}
    $RNG.seed(h)$\\
    $S \leftarrow \texttt{shuffled}(N, RNG)$\\
    $S \leftarrow S.slice(0,\ns)$\\
	\Return $S \;$\\
\end{algorithm}

\begin{algorithm}[t]
\SetAlgoRefName{{\tt checkMessage}}
\caption{Check the message received from node $u$}
\label{algo:check message interaction}
\DontPrintSemicolon
\KwIn{
	$u$, the sender of the request \\
	$h$, difficulty of the PoI\\
	$d$, the dependency (hash of last block of the blockchain) \\
	$m$, the message (root of the merkle tree of the new block) \\
}
\KwOut{
	whether to accept or not the request
}
	\eIf{$d$ is the hash of the latest block of one of the longest branches}{
        \If{Received[$(u,d)$] exists and is not equal to $m$}{
            \penalties($u$) \;
            \textbf{Reply} false \;
        }
        Received[$(u,d)$] = $m$ \;
        \textbf{Reply} true
    }
    {
        \If{unknown $d$}{
            \textbf{Ask block} $d$
        }
    }
\end{algorithm}

\begin{figure}
    \centering
\begin{tikzpicture}[scale=0.5]
\foreach \x in {0,...,15}
\node at (\x+0.5,20.5) {\scriptsize \x};
\foreach \x in {0,...,16}
\draw[thick, blue] (\x,20) -- (\x,21);
\node[thick] (bit1) at (-1,20.5) {\scriptsize bytes};

\def\x{17}
\draw [thick] (-1, \x+2) -- (-0.1,\x+2);
\draw [<->, thick] (-0.5, \x+1.9) -- (-0.5,\x+0.1);
\fill[white] (-0.8,\x+1.2) rectangle (-0.1,\x+0.8);
\node at (-0.95,\x+1) {\tiny 32 bytes};
\draw [thick] (-1, \x) -- (-0.1, \x);

\def\x{15}
\draw [thick] (-1, \x+2) -- (-0.1,\x+2);
\draw [<->, thick] (-0.5, \x+1.9) -- (-0.5,\x+0.1);
\fill[white] (-0.8,\x+1.2) rectangle (-0.1,\x+0.8);
\node at (-0.95,\x+1) {\tiny 32 bytes};
\draw [thick] (-1, \x) -- (-0.1, \x);

\def\x{13}
\draw [thick] (-1, \x+2) -- (-0.1,\x+2);
\draw [<->, thick] (-0.5, \x+1.9) -- (-0.5,\x+0.1);
\fill[white] (-0.8,\x+1.2) rectangle (-0.1,\x+0.8);
\node at (-0.95,\x+1) {\tiny 32 bytes};
\draw [thick] (-1, \x) -- (-0.1, \x);

\filldraw[thick,draw=black, fill=white] (0,20) rectangle (4,19); \node (mode) at (2,19.5) {\scriptsize Version};
\filldraw[thick,draw=black, fill=white] (4,20) rectangle (8,19); \node (mode) at (6,19.5) {\scriptsize Time};
\filldraw[thick,draw=black, fill=white] (8,20) rectangle (12,19); \node (mode) at (10,19.5) {\scriptsize Difficulty};
\filldraw[thick,draw=black, fill=white] (12,20) rectangle (16,19); \node (mode) at (14,19.5) {\scriptsize Favorite Cheese};
\filldraw[thick,draw=black, fill=white] (0,19) rectangle (16,17); \node (mode) at (8,18.5) {\scriptsize Previous block header hash};
\filldraw[thick,draw=black, fill=white] (0,17) rectangle (16,15); \node (mode) at (8,16.5) {\scriptsize Merkle root hash};
\filldraw[thick,draw=black, fill=white] (0,15) rectangle (16,13); \node (mode) at (8,14.5) {\scriptsize Proof hash};
\end{tikzpicture}
    \caption{Definition of a Block, for our PoI based Blockchain}
    \label{fig:block}
\end{figure}
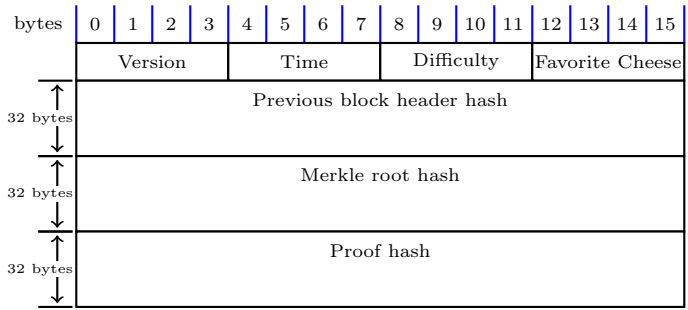

\end{document}